\documentclass[12pt]{article}
\usepackage{graphicx}
\usepackage{amsmath}

\newtheorem{theorem}{Theorem}[section]
\newtheorem{lemma}{Lemma}[section]

\newtheorem{definition}{Definition}[section]
\newtheorem{example}{Example}[section]

\def\addcontentsline#1#2#3{\relax}

\long\outer\def\demo#1. #2\par{\medbreak\noindent {\bf#1.\enspace}
        {\rm#2}\par\ifdim\lastskip<\medskipamount\removelastskip
        \penalty55\medskip\fi}
\newcommand{\ben}{\begin{enumerate}}
\newcommand{\een}{\end{enumerate}}
\def\bdemo#1. #2\par{\medbreak\noindent {\bf#1.\enspace}{\rm#2}\par}
\def\edemo{\ifdim\lastskip<\medskipamount\removelastskip\penalty55\medskip\fi}

\def\p{{\partial}}

\def\0{{\bf 0}}
\def\1{\large{\bf\large 1}}

\def\b{{\bf b}}

\def\l{\textbf{\emph{l}}}
\def\n{{\bf n}}
\def\p{{\bf p}}

\def\v{{\bf v}}

\def\q{{\bf q}}

\def\x{{\bf x}}
\def\y{{\bf y}}
\def\z{{\bf z}}
\def\N{{\bf N}}
\def\Q{{\bf Q}}
\def\R{{\bf R}}
\def\Z{{\bf Z}}

\def\Vol{{\rm Vol}}

\def\mod{\mathop{\rm mod}\nolimits}

\def\longto{\mathop{\longrightarrow}\limits}
\def\bfa{\mathop{a\kern-.5em{a}\kern-.5em{a}}\nolimits}
\def\bfb{\mathop{b\kern-.47em{b}\kern-.47em{b}}\nolimits}
\def\bfx{\mathop{x\kern-.6em{x}\kern-.6em{x}}\nolimits}
\def\bfalpha{\mathop{\alpha\kern-.6em{\alpha}}\nolimits}
\def\bftheta{\mathop{\theta\kern-.53em{\theta}}\nolimits}
\begin{document}
\baselineskip 18pt
\title{Discrete charging of metallic grains: Statistics of addition spectra}
\author{Yshai Avishai\footnote{%
Departments of Physics, Ben-Gurion University of the Negev,
Beer-Sheva 84105, Israel},\texttt{ }
 Daniel Berend\footnote{%
Departments of Mathematics and Computer Science, Ben-Gurion
University of the Negev, Beer-Sheva 84105, Israel}   \texttt{ }and
Luba
Bromberg\footnote{%
Department of Mathematics, Ben-Gurion University of the Negev,
Beer-Sheva 84105, Israel}
 }
\date{}
 \maketitle
 \begin{abstract}
We analyze the statistics of electrostatic energies (and their
differences) for a quantum dot system composed of a finite number
$K$ of electron islands (metallic grains) with random 
capacitance-inductance matrix $C$, for
which the total charge is discrete, $Q=Ne$ (where $e$ is the
charge of an electron and $N$ is an integer). 
The analysis is based on a generalized charging model, where
the electrons are distributed among the grains
such that the electrostatic energy $E(N)$ is minimal. Its second
difference (inverse compressibility) $\chi_{N}=E(N+1)-2
E(N)+E(N-1)$ represents the spacing between 
adjacent Coulomb blockade peaks
appearing when the conductance of the quantum dot is plotted
against gate voltage. The statistics of this quantity has been the
focus of experimental and theoretical investigations during the
last two decades.  We provide an algorithm for calculating the
distribution function corresponding to $\chi_{N}$ and show that
this function is piecewise polynomial.
\end{abstract}
\section{Introduction}
The physics exposed in the addition spectra of quantum dots is
rather rich, and hence its investigation is at the focus of both
experimental and theoretical studies. After
the origin of Coulomb-blockade peaks has been elucidated,
investigation is directed toward more subtle questions such as
their heights, widths, and spacings. The underlying physics is
related to the ground-state energy, chemical potential, and
inverse compressibility of quantum dots composed of a few
metallic electron islands coupled capacitively and inductively to
each other.  

The present work
concentrates on the distribution of spacings between
Coulomb-blockade peaks in large semiconductor quantum dots. In particular, we are interested in fluctuations of
these quantities with the number $N$ of electrons on the dot. The
main problem can be stated as follows: According to the simplest
picture (charging model), 
in which the quantum dot is regarded as a single-electron
island whose coupling with the leads is through its capacitance
$C$, the total potential energy of a quantum dot with $N$
electrons and charge $Q=Ne$ is $Q^{2}/2C-V_{g}Q,$  where $V_{g}$
is the gate voltage and $e$ is the electron charge. The position
of the $N$-th Coulomb-blockade peak occurs at a gate voltage
$V_{g}=\frac {N e^{2}}{C}$. This peak position is
 then a linear function of $N$,
and therefore the spacing between 
two adjacent peaks should be a constant $e^{2}/C$ ,
independent of $N$. This is not always confirmed experimentally.
The situation is even more intriguing if the quantum dot is 
large and might contain more than a single electron puddle. As indicated in a series of recent experiments
\cite{Ashoori}, the spacing between adjacent 
Coulomb blockade peaks occasionally vanishes; namely, Coulomb
blockade peaks tend to bunch. The problem is therefore to explain
why the results predicted from a simple charging model deviate
substantially from the experimental observation.

In \cite{ABB1}, a generalized charging model has been tested,
where it is assumed that the large dot used in the experiments
\cite{Ashoori} could be divided into a set of potential wells
(metallic grains) with random capacitances and 
random mutual inductances.
This casts the question of Coulomb blockade peak spacing
distribution into the problem of elucidating the statistics of the addition
spectrum of a relatively simple physical system. It consists of
$K$ metallic grains (or capacitors), such that the number of
electrons on the $i$-th grain is $n_{i}$ ($i=1,2,\ldots,K$),
the total number of electrons being $N$. The charging model for
such a system (at zero temperature)
is based on the assumption that the distribution of
electrons among the grains is determined by requiring that the
electrostatic energy $E(N)$ of a dot containing $N$ electrons is
minimal. It is useful at this point to recall the basic facts
pertaining to the energy of the electrostatic field of conductors
\cite{Landau}. The electrostatic energy of the system is a
bilinear form in the numbers $n_{i}$. This form is given by a $K
\times K$ matrix $W =\frac {1} {2} C^{-1}$. Here $C$ 
(matrix elements $c_{ij}; \ \ i,j=1,2...K)$ is a
positive-definite symmetric matrix of capacitance and inductance
coefficients. Physically, the matrix $C$ has positive diagonal
entries and {\it negative} (more precisely, non-positive) 
non-diagonal entries \cite{Landau},
\begin{eqnarray}
&& c_{ij}=c_{ji}, \ \ \ c_{ii}>0, \ \ \  c_{ij}\leq0 \ \  (i \ne
j). \label{CS}
\end{eqnarray}
On the other hand, all the elements of $C^{-1}$ are non-negative. 
More precisely, with $w_{ij}; \ \ i,j=1,2...K$ the elements of the matrix 
$W =\frac {1} {2} C^{-1}$, one has,
\begin{eqnarray}
&& w_{ii}>0, \ \ \  w_{ij}\geq0 \ \  (i \ne
j). \label{Cinv}
\end{eqnarray}
 The off-diagonal entries $c_{ij}, i \ne j$,
 decay as an inverse power of the distance between the
grains, while the diagonal entries $c_{ii}$ are proportional to
the geometrical size of the grains. 
The notion of randomness enters when we recall that
 experimentally,  the sizes of the grains, as well as
 the distances between them, are random quantities.
This means that the elements of the matrix $C$ are random
numbers (subject, of course, to the required symmetries
(\ref{CS})). The spacing between Coulomb blockade peaks is equal
to the second difference of the ground state energy. In other
words, the distribution of spacing peaks is determined by the
statistics of the inverse compressibility,
\begin{eqnarray}
\chi_{N} \equiv E(N+1)-2 E(N)+E(N-1).
\end{eqnarray}
When two Coulomb blockade peaks coincide, the second difference
$\chi_{N}$ vanishes. Note that, on the average (and on a large
scale), the energy $E(N)$ grows quadratically with~$N$.
Therefore, one would expect the second difference to stay finite
and $N$ independent. However, there is no simple relation such as
$E(N)=a N +b N^{2}$. The deviation of $E(N)$ from exact quadratic
behavior makes its second difference $\chi_{N}$ non-constant, and
a fluctuating quantity. It is precisely these fluctuations which
we intend to study. As we shall see, the fact that electron charge
is quantized makes this task non-trivial.

 Having explained the physical motivation, we then
 pose the mathematical problem: what is the distribution $p(\chi)$ of
 inverse compressibility  for a given system of
  metallic grains with random capacitive matrix $C$?
    As a crude approximation it was assumed in \cite{Berend} that the
metallic grains are indeed very far apart, and the matrix $C$ is
nearly diagonal, its $K$ diagonal elements (capacitances) being
random numbers. The energy of the system in the diagonal case is
given by
\begin{eqnarray}
E(N)=\min \sum_{i=1}^{K} \frac {1} {2 c_{i}} n_{i}^{2},\qquad
\mbox {(subject to $\sum_{i=1}^{K} n_{i}=N).$} \label{eq_EN}
\end{eqnarray}

The minimum in (\ref{eq_EN}) is taken over all possible partitions
$(n_i)_{i=1}^{K}$ of $N$.
 It was first proved that the minimum problem (\ref{eq_EN}) has the
following convenient feature: If $n_1,n_2,\ldots,n_{K}$ are the
argument values bringing $E(N)$ to its minimal value for some $N$,
then the minimum for $N+1$ is obtained by retaining all $n_i$'s,
except for one which is increased by~1. This allowed an exact
determination of the distribution function according to which the
sequence~$\chi_{N}$ is distributed. For a random set of capacitors
($c_{1}, c_{2},...,c_{K}$ -- random numbers with probability
distribution $P(c_{1}, c_{2},...,c_{K})$), the distribution of the
inverse compressibility $F(\chi)$ was calculated in \cite{Berend}.

Our next goal is to study this problem for general
positive-definite matrices~$C$. The problem turns out to be quite
harder. To begin with, it is no longer true that the optimal
solution for $N+1$ is obtained in a simple manner from that
for~$N$. That is, for each $N$ we need to re-distribute the~$N$
electrons between the grains, and it may well happen that,
although the $n_i$'s grow in general with $N$, some of them will
actually decrease infinitely often as $N$ increases by~1 each
time. Namely, there will exist infinitely many values of~$N$ for
which the optimal value of some $n_i$ decreases as~$N$ grows
to~$N+1$. Thus, the problem entails new behavioral patterns with
respect to the diagonal case.

Our main result in this paper is an algorithm for calculating the
distribution function corresponding to $(\chi_{N})$. Moreover, we
show that this function is piecewise polynomial. We state the
result in Section 2. Section 3 is a short digression, discussing a
few notions which arise in the proof. The proof of the main
theorem is given in Section 4.
\section{The Main Results}
\def\Vol{{\rm Vol}}
\def\deg{{\rm deg}}
\bigskip
Mathematically, our problem is as follows. Let
$C=(c_{ij})_{i,j=1}^K$ be a positive-definite matrix, with
positive diagonal elements and non-positive off-diagonal elements.
Assume that the sum of elements in every row of $C$ is positive
and that all entries of $\frac{1}{2}C^{-1}=W=(w_{ij})_{i,j=1}^K$
are non-negative. Put
\begin{equation}\label{constrained}
E(N)= \min\left\{\sum_{i,j=1}^K w_{ij} n_i n_j:
  \,n_i\in\Z_+,\,\sum_{i=1}^K n_i=N\,\right\},\qquad N\in\N,
\end{equation}
where $\N$ is the set of positive integers and $\Z_+=\N\cup\{0\}$.
We want to understand the statistical behavior of the sequence
$E(N)$, and in particular that of the second difference sequence
\begin{equation}\label{chi0}\chi_N=E(N+1)-2 E(N)+E(N-1).\end{equation}
To formulate our main result, we need a few definitions and
notations.
\begin{definition}\label{def1}
\rm Let $(x_{n})_{n=1}^{\infty}$ be a sequence of real numbers and
$F$ a distribution function. The sequence $(x_{n})$ is
\emph{asymptotically $F$-distributed} if
$$ {\frac{|\{1\leq n\leq M: x_{n}\leq x\}|}{M}}\longto_{M\to\infty} F(x)$$
for every continuity point $x$ of $F$ (where $|S|$ denotes the
cardinality of a finite set $S$).
\end{definition}

The definition almost coincides with \cite[p.53, Def. 7.1]{KN},
except that there the sequence $(x_{n})$ is considered only modulo
1. Note that a sequence need not be asymptotically $F$-distributed
for some $F$, as the following example shows.
\begin{example} The sequence of numbers   $$ 0, \ \
\underbrace{1,...,1,}_{10}\ \underbrace{0,...,0,}_{10^{2}}\ \
\underbrace{1,...,1,}_{10^{3}}...$$ is not asymptotically
$F$-distributed for any $F$.
\end{example}

A stronger notion is obtained when we require not only long
initial block of the sequence to behave approximately according to
$F$, but rather require any long block to behave so. This leads to
the following definition (\cite[p.40, Def. 5.1]{KN} and
\cite[p.200, Def. 3.2]{KN}.)
\begin{definition} \rm In the setup of Definition \ref{def1}, $(x_{n})$ is
\emph{asymptotically well $F$-distributed} if
$$\frac{\left|\{L< n\le M:x_n\le x\}\right|} {M-L}\longto_{M-L\to\infty} F(x)$$
for every continuity point $x$ of $F$.
\end{definition}

The following example demonstrates that the property of asymptotic
 well $F$-distribution is indeed strictly stronger than that of
 asymptotic $F$-distribution.
\begin{example} The sequence of numbers
  $$\underbrace{0}_{1},\underbrace{1}_{1},\underbrace{0,0}_{2},\underbrace{1,1}_{2},
  \underbrace{0,0,0}_{3},\underbrace{1,1,1}_{3},...$$
  is asymptotically $F$-distributed, where $F$ is the distribution function
\begin{equation*}
F(x)= \left\{\begin{aligned}
&0,\qquad x<0,\\
      &\frac{1}{2}, \qquad 0\leq x< 1,\\
      &1, \qquad x\geq1,
\end{aligned}
\right.
 \end{equation*}

 \noindent but it is not asymptotically well $F$-distributed.
\end{example}

\begin{definition}\rm A function $g:\mathbf{R}\rightarrow
\mathbf{R}$ is \emph{piecewise polynomial} if there exist
intervals (finite or infinite) $I_{j}\subseteq \mathbf{R}$ and
polynomials $Q_{j}$, $1\leq j \leq m$, such that $$g(x)=Q_{j}(x),
\qquad x \in I_{j}, \ 1\leq j \leq m.$$ The \emph{degree} of $g$
is $ \underset{1\leq j\leq m}{\max} \deg~Q_{j}$.
\end{definition}

Returning to our problem, let  $b_{i}=\sum_{j=1}^{K}c_{ij}, 1\leq
i\leq K$,
 be the row sums of the matrix $C$. Since the matrix $C$ is
 random, in the generic case the numbers $b_{1},b_{2},...,b_{K}$
 are linearly independent over the rationals. (That is, considered
 as vectors in the space $\mathbf{R}$ over the field of rational
 numbers $\mathbf{Q}$, they are independent.)

  Now we can formulate our main result.
\begin{theorem}  \label{th1} \rm

Let $C$ be a positive-definite symmetric matrix, with positive row
sums $b_{1},b_{2},...,b_{K}$ and let $W=\frac{1}{2}C^{-1}$.
Suppose that $b_{1},b_{2},...,b_{K}$ are linearly independent over
the rationals. Then the sequence $(\chi_{N})_{N=1}^{\infty}$ of
the second differences, defined via (\ref{constrained}) and
(\ref{chi0}), is asymptotically well $F$-distributed, where $F$ is
a continuous piecewise polynomial function of degree at most
$K-1$, which can be effectively computed.
\end{theorem}

As mentioned in the introduction, a phenomenon which occurs in the
general case dealt with here, but not in the special case of
diagonal matrices~$C$, is that, as we pass from $N$ to $N+1$,
there may be re-distribution of the $n_{i}$'s in the optimal
solution. The following example is to that effect.
\begin{example}
Let
\begin{equation*}
C=\left(
\begin{array}{ccc}
2 & 0 & -1 \\
0 & 2 & -1 \\
-1 & -1 & \ \ 3
\end{array}
\right),\qquad W=\frac{1}{2}C^{-1}=\frac{1}{16}\left(
\begin{array}{ccc}
5 & 1 & 2 \\
1 & 5 & 2 \\
2 & 2 & 4
\end{array}
\right).
\end{equation*}
Then:
$$E(N)=\frac{1}{16}\min_{n_{1}+n_{2}+n_{3}=N, n_{i}\geq 0}(5n_{1}^2+2n_{1}n_{2}+4n_{1}n_{3}+5n_{2}^2+4n_{2}n_{3}+4n_{3}^2).$$
Using the techniques in the beginning of Section 4, it is easy to
verify that the optimal values of $n_{1},n_{2},n_{3}$ are given by
\begin{equation*}(n_{1},n_{2},n_{3})=\left\{
\begin{array}{ll}
(~ \frac{N}{3},\ \ \ \frac{N}{3},\ \ \ \frac{N}{3}~), & N\equiv0(\mod 3), \\
(\frac{N-1}{3},\frac{N-1}{3},\frac{N+2}{3}), & N\equiv1(\mod 3), \\
(\frac{N+1}{3},\frac{N+1}{3},\frac{N-2}{3}), & N\equiv2(\mod 3).
\end{array}\right.
\end{equation*}
Thus,for any non-negative integer $K$, when passing from $N=3k+1$
to $N=3k+2$, the value of $n_{3}$ at the optimal point decreases
from $k+1$ to $k$.
\end{example}
\section{Uniform Distribution Modulo 1}
In this section we briefly discuss the notion of uniform
distribution modulo~1 and recall a few related results, which will
be needed in the proof of the Theorem 2.1.
\begin{definition} \rm
A sequence $(x_n)_{n=1}^\infty$ of real numbers is {\it uniformly
distributed modulo}~1 if
$$\frac{\left|\{1\le n\le N:a\le \{x_n\}<b\}\right|}{N}\longto_{N\to\infty}
        b-a,\qquad 0\le a<b\le 1\;,$$
where $\{t\}$ is the fractional part of a real number $t$
(\cite[p.1, Def. 1.1]{KN}).

 In terms of Definition \ref{def1}, $(x_n)$ is uniformly distributed modulo~1
if and only if the sequence $(\{x_n\})$ of fractional parts is
$F$-distributed, where $F$ is the distribution function of the
uniform distribution on $[0,1]$:

\begin{equation}\label{f}
F(x)= \left\{\begin{aligned}
&0,\qquad x<0,\\
      &x,\qquad 0\leq x\leq 1,\\
      &1,\qquad x>1.
\end{aligned}
\right.
 \end{equation}
\end{definition}

 The notion of uniform distribution modulo 1 has a
multi-dimensional analogue. A sequence
$(\mathbf{x}_n)_{n=1}^\infty$ in $\R^s$ is {\it uniformly
distributed modulo~1 in} $\R^s$ if
$$\frac{\left|\{1\le n\le N:\mathbf{a}\le \{\mathbf{x}_n\}<\mathbf{b}\}\right|}{N}\longto_{N\to\infty}
        \prod_{i=1}^s (b_i-a_i),\qquad \0\le \mathbf{a}<\mathbf{b} \le \1,$$
where $\0=(0,0,\ldots,0)\in\R^s,\
\mathbf{a}=(a_1,a_2,\ldots,a_s)$, and so forth, and inequalities
between vectors in $\R^s$ are to be understood component-wise
(\cite[p.47, Def. 6.1]{KN}).

The notion of uniform distribution modulo 1, both in the
1-dimensional and the multi-dimensional cases, has a stronger
version, whereby the required property holds not only along
initial blocks of the sequence, but along any blocks of larger and
larger lengths (\cite[p.40, Def. 5.1]{KN}. A sequence satisfying
this stronger property is \it {well distributed modulo 1}. \rm
Obviously, well distribution modulo 1 is equivalent in the
1-dimensional case to $F$-distribution of the sequence of
fractional parts for the function $F$ given by (\ref{f}).
 A basic example of a sequence which is
uniformly distributed modulo~1 is $(n\alpha)_{n=1}^\infty$, where
$\alpha$ is an arbitrary irrational \cite[p.8, Def 2.1]{KN}. In
the multi-dimensional case, the sequence
$(n\alpha_1,n\alpha_2,\ldots,n\alpha_s)_{n=1}^{\infty}$ is
uniformly distributed modulo~1 in $\R^s$ if and only if the
numbers $1,\alpha_1,\alpha_2,\ldots,\alpha_s$ are linearly
independent over $\Q$ \cite[pp. 48-49]{KN}. Moreover, for these
sequence, well distribution is equivalent to uniform distribution.

 Recall that the
\emph{density} of a set $A\subseteq \textbf{N}$ is given by
$$ D(A)=\lim_{M\rightarrow\infty}\frac{|A\cap[1,M]|}{M}$$
if the limits exists. If, moreover, the limit
$$ BD(A)=\lim_{M-L\rightarrow\infty}\frac{|A\cap(L,M]|}{M}$$
exists, then it is called the \emph{Banach density} of $A$.

We can rephrase the definition of uniform distribution modulo~1
using the notion of density of a set. Namely,
$(x_{n})_{n=1}^{\infty}$ is uniformly distributed modulo~1 if for
every interval $I \subseteq [0,1)$ we have
\begin{equation}\label{D}
D(\{n:\{x_n\} \in I\})=|I|,
\end{equation}
 where $|I|$ denotes the length of
$I$. Similarly, $(x_{n})_{n=1}^{\infty}$ is well distributed
modulo 1 if (\ref{D}) continues to hold when the density of the
left-hand side is replaced by Banach density.

\section{Proof of Theorem 2.1}

To avoid complicated notation, shall prove in Theorem 2.1 only
that $(x_{N})_{N=1}^{\infty}$ is asymptotically $F$-distributed,
and not that it is asymptotically well $F$-distributed. As will be
seen in the proof, our result depends on the fact that the
sequence $(\{Nb_{1}\},\{Nb_{2}\},...,\{Nb_{K-1}\})$ is uniformly
distributed modulo 1 in $\textbf{R}^{K-1}$. Since this sequence is
actually well distributed modulo 1, the same proof shows that
$(x_{N})_{N=1}^{\infty}$ is actually well $F$-distributed.

 \noindent Along with the sequence $E(N)$ from
(\ref{constrained}), it is very useful to consider the sequence
$E_1(N)$, defined by
\begin{equation}\label{unconstrained}
E_1(N)= \min\left\{\sum_{i,j=1}^K w_{ij} x_i x_j:
  \,x_i\in\R,\,\sum_{i=1}^K x_i=N\,\right\},\qquad N\in\N.
\end{equation}
Obviously, $E_1(N)\le E(N)$ for each $N$. We shall refer to the
minimum problems on the right hand side of (\ref{constrained}) and
of (\ref{unconstrained}) as the {\it constrained problem} and the
{\it unconstrained problem}, respectively.

Denote by $\mathbf{e}$ the column $K$-vector with all entries 1.

 \begin{lemma}\label{lem1}  The unique minimum of the unconstrained problem is
$$\x_0=\frac{N}{\sum_{i,j=1}^K c_{ij}}\cdot C\,\mathbf{e}\,.$$ and
the corresponding unconstrained minimum is
$E_{1}(N)=\frac{N^{2}}{2\sum_{i,j=1}^K c_{ij}}.$
 \end{lemma}
Due to our assumption regarding the positivity of the row sums of
$C$, all components of $\mathbf{x}_0$ are positive. Multiplying
all entries of $C$ by any constant $c>0$ we obtain an equivalent
problem. Taking  $c=(  \sum_{i,j=1}^K c_{ij} ) ^{-1}$, \bf{we
shall henceforth assume that} \rm $ \sum_{i,j=1}^K c_{ij} =1$. In
particular, denoting $\b=(b_{1},b_{2},...b_{K})^t$, we have
 \begin{equation}\label{b_sol}
 \x_0=NC\mathbf{e}=N \b
 \end{equation}\ and
 \begin{equation}\label{sol}E_{1}(N)=\frac{N^2}{2}.
 \end{equation}

 \noindent\bdemo Proof of Lemma \ref{lem1} \rm  Let $\x\ne\x_0$ be any
feasible solution of the unconstrained problem. Putting
$a=\displaystyle{\frac{N}{\sum_{i,j=1}^K c_{ij}}}$ and
$\y=\x-\x_0$, we obtain
$$\begin{array}{ll}
\x^t W\x &=(\x_0+\y)^t W(\x_0+\y)=\x_0^t W\x + 2\x_0^t W\y + \y^t
W\y\cr
         &=\x_0^t W\x_0 + a\1^t C C^{-1} \y + \y^t W\y\cr
         &=\x_0^t W\x_0 + a\1^t \y + \y^t W\y=\x_0^t W\x_0
           + \y^t W\y>\x_0^t W\x_0\,.
\end{array}$$
Consequently, $$E_{1}(N)=\frac{N}{\sum_{i,j=1}^K
c_{ij}}\mathbf{e}^tC W \frac{N}{\sum_{i,j=1}^K
c_{ij}}C\mathbf{e}=\frac{N^2}{2\sum_{i,j=1}^K c_{ij}}.$$ \edemo
 \begin{lemma}\label{lem2}
  $E_1(N)\ge E(N)-\sum_{i,j=1}^K w_{ij}$ for
every $N$.
 \end{lemma}

\noindent\textbf{Proof} Let $\x_0=(x_{01},x_{02},\ldots,x_{0K})$
be the minimum point of the unconstrained problem. Let
$r=\sum_{i=1}^K \{x_{0i}\}$ be the sum of fractional parts of all
coordinates of $\x_0$. Obviously, $r$ is an integer, $0\le r<K$.
Let $i_1,i_2,\ldots,i_K$ be all integers between $1$ and $K$,
ordered so that $\{x_{0,i_1}\}\le \{x_{0,i_2}\}\le\ldots\le
\{x_{0,i_K}\}$ (where ties are resolved arbitrarily). Consider the
vector $\n=(n_1,n_2,\ldots,n_K)$ defined by
$$n_i=\left\{\begin{array}{lll}
  & [x_{0,i}],\qquad & i=i_1,i_2,\ldots,i_{K-r},\cr
  & [x_{0,i}]+1,\qquad & {\rm otherwise\,.}\cr
  \end{array}
  \right.$$
As mentioned in Lemma \ref{lem1}, all $x_{0,i}$'s are positive,
and hence $\n$ is a feasible solution of the constrained problem.
Set $\y=\n-\x_0$. Since all coordinates of $\y$ lie in the
interval $(-1,1)$, as in the proof of Lemma \ref{lem1} we have
\begin{equation}\label{connect}
E(N)\le  n^t W\n
      = x_0^t W\x_0 + y^t W\y
      \le E_1(N) + \sum_{i,j=1}^K
      w_{ij}\,\end{equation}
which proves the lemma.
  \begin{lemma}\label{lem3}
   There exists an effective
constant $\Delta=\Delta(C)$ such that, for every $N$, the distance
between the solution of the constrained problem and that of the
unconstrained problem does not exceed $\Delta$.
 \end{lemma}

\noindent\textbf{ Proof} Write $W=P^{-1}DP$, where $P$ is
orthogonal and $D$ diagonal. Let $M$ be an upper bound on the
eigenvalues of $C$ (for example, the $L^\infty$-norm $\max_{1\le
i\le K} \sum_{j=1}^K |c_{ij}|$ of $C$). Then $M^{-1}$ is a lower
bound for the eigenvalues of $W$, namely for the diagonal entries
of $D$. Let $F$ be the diagonal matrix with positive diagonal
entries and $F^2=D$. Obviously, $\|F\z\|_2\ge M^{-1/2}\|\z\|_2$
for every $\z\in\R^K$. Then for every $\y\in\R^K$ we have
$$\y^t W \y=\y^t P^{-1} FFP\y=\|FP\y\|_2^2\ge
\|P\y\|_2^2/M=\|\y\|_2^2/M\,.$$ Now let $\x_0$ and $\n=\x_0+\y$ be
minimum points of the unconstrained problem and of the constrained
problem, respectively. Then
$$E(N)=\n^t W \n=\x_0^t W\x_0 + \y^t W \y,$$
which implies by Lemma \ref{lem2} that $\y^t W \y\le
\sum_{i,j=1}^K w_{ij}$. Thus
$$\|\y\|_2^2/M\le \sum_{i,j=1}^K w_{ij},$$
which yields the conclusion of the lemma with
$$\Delta=\sqrt{M\sum_{i,j=1}^K w_{ij}}.$$
\textbf{ Proof of Theorem 2.1.} Lemmas \ref{lem1}-\ref{lem3}
provide a simple algorithm for calculating $E(N)$ for each $N$ in
constant time. Namely, we find the point $\x_0$ yielding the
optimal $E_1(N)$ according to Lemma \ref{lem1}, calculate the
value of $\n^t W\n$ for all integral points $\n$, with coordinate
sum $N$, within distance $\Delta$ from $\x_0$, and take the best
of them. If the optimal point turns out to be $\n=\x_0+\y$, we
shall refer to $\y$ as {\it the correction vector}. We have
$\y=\n-\x_{0}=\l-\{\x_0\}$, where $\{\x_0\}$ denotes the vector of
fractional parts of the coordinates of $\x_0$ and $\l$ belongs to
some finite effective set $L$ of integer vectors. Since the sum of
coordinates of the correction vector is always 0, the sum of
coordinates of $\l$ must equal that of $\{\x_0\}$. Thus, $L$
consists of all integer vectors $\l$, for which the vector
$\l-\{\x_0\}$ is of norm not exceeding the bound in Lemma
\ref{lem3} and its coordinates sum vanishes. To emphasize the
dependence of $L$ on $\x_{0}$, we shall sometimes write
$L(\x_{0})$ instead of $L$.

Now when choosing the optimal $\l$ out of $L$, we first notice
that, among any two candidates $\l_1$ and $\l_2$, the former will
be better (or equal) than the latter if and only if
$$(\x_0+\l_1-\{\x_0\})^t W (\x_0+\l_1-\{\x_0\}) \le
  (\x_0+\l_2-\{\x_0\})^t W (\x_0+\l_2-\{\x_0\})\,.$$
This inequality is easily seen to be equivalent to
$$2(\l_2-\l_1)^t W \{\x_0\} \le \l_2^t W \l_2 -\l_1^t W \l_1\,.$$
Consequently, $\l$ is the optimal choice if and only if
\begin{equation}\label{for1}
2(\l'-\l)^t W \{\x_0\} \le (\l')^t W \l' -\l^t W \l,\qquad \l'\in
L\,.
\end{equation}
To study  the second differences
\begin{equation*}
\chi_{N}= E(N+1)-2E(N)+E(N-1),
\end{equation*}
we shall write each term on the right-hand side in the form
$E_{1}(N+j)+d_{j}$ for an appropriate $d_{j}$. In fact, as in
(\ref{connect}), denoting by $\y_{1}, \y_{2}$ and $\y_{3}$ the
correction vectors for $N-1,N$ and $N+1$, respectively, we have:
\begin{equation}\label{chi}
\chi_{N}=
E_{1}(N+1)-2E_{1}(N)+E_{1}(N-1)+\y_{3}^tW\y_{3}-2\y_{2}^tW\y_{2}+\y_{1}^tW\y_{1}.
\end{equation}
By (\ref{sol}):
\begin{equation}
\chi_{N}= 1+\y_{3}^tW{2}\y_{3}-2\y_{2}^tW\y_{2}+\y_{1}^tW\y_{1}.
\end{equation}
 Let be $\x_{0}, \x'_{0}, \x''_{0}$  the
points yielding the optimal values of $E_{1}(N-1)$, $E_{1}(N)$,
$E_{1}(N+1)$, respectively. In view of (\ref{b_sol}):
\begin{equation}\label{connec}\x'_{0}=\x_{0}+\b,\qquad
\x''_{0}=\x_{0}+2\b.
\end{equation}
For appropriate integer vectors $\p \in L(\x_{0}), \p' \in
L(\x'_{0}),\p'' \in L(\x''_{0})$:
 \begin{equation}\label{vec}\y_{1}=\p-\{\x_{0}\},\
\y_{2}=\p'-\{\x'_{0}\},\ \y_{3}=\p''-\{\x''_{0}\}.
\end{equation}
 The vectors  $\p, \p', \p''$ are determined by the system of
 inequalities:
\begin{equation}\label{doman1}
\left\{\begin{array}{lll} &2(\l-\p)^{t}W\{\x_{0}\}\leq
\l^{t}W\l-\p^{t}W\p, \qquad &\l \in
L(\x_{0}),\\
&2(\l-\p')^{t}W\{\x'_{0}\}\leq
 \l^{t}W\l-(\p')^{t}W\p', \qquad &\l \in L(\x'_{0}),\\
&2( \l-\p'')^{t}W\{\x''_{0}\}\leq \l^{t}W\l-(\p'')^{t}W\p'',
\qquad &\l \in L(\x''_{0}).
\end{array} \right.
\end{equation}
Due to (\ref{connec}), it is natural to try to rewrite
(\ref{doman1}) in terms of $\{\x_{0}\}$ without referring to
$\{\x'_{0}\}$ and $\{\x''_{0}\}$. Divide the $K$-dimensional torus
$\textbf{T}^{K}$, which we identify with $ [0,1)^{K}$, according
to the vector $\b$, as follows.

The $i$-th coordinate $\{x'_{0i}\}$ of $\{x'_{0}\}$ may be either
 $\{x_{0i}\}+b_{i}$ or $\{x_{0i}\}+b_{i}-1$, depending on
 whether $\{x'_{0i}\}+b_{i}$ is smaller than $1$ or not,
 respectively. Similarly, $\{x''_{0i}\}$ may assume one of the three
 values $\{x_{0i}\}+2b_{i}-c$, where $c=0,1,2.$ Divide the
 circle $\textbf{T}$ into three disjoint intervals (actually arcs), on each of which
 both $\{x'_{0i}\}$ and $\{x''_{0i}\}$ assume the same form in
 terms of $\{x_{0i}\}.$ We have to distinguish between two cases:

 \noindent 1) If $ b_{i}\leq \frac{1}{2},$ write
\begin{equation}\label{int1}
\textbf{T}=[0,1-2b_{i})\cup[1-2b_{i},1-b_{i})\cup[1-b_{i},1).
\end{equation}
If $\{x_{0i}\}$ belongs to the first interval on the right-hand,
then
\begin{equation*}
\{x_{0i}'\} =\{x_{0i}\}+b_{i},\qquad
\{x_{0i}''\}=\{x_{0i}\}+2b_{i},
\end{equation*}
if it belongs to the second
\begin{equation*}
\{x_{0i}'\} =\{x_{0i}\}+b_{i},\qquad
\{x_{0i}''\}=\{x_{0i}\}+2b_{i}-1,
\end{equation*}
and if it belongs to the third
\begin{equation*}
 \{x_{0i}'\} =\{x_{0i}\}+b_{i}-1,\qquad
\{x_{0i}''\}=\{x_{0i}\}+2b_{i}-1.
\end{equation*}
 \noindent 2) If $ b_{i}> \frac{1}{2},$ write
\begin{equation}\label{int2}
\textbf{T}=[0,1-b_{i})\cup[1-b_{i},2-2b_{i})\cup[2-2b_{i},1).
\end{equation}
This time, depending on the interval on the right-hand side
containing $\{x_{0i}\}$, we have either
\begin{equation*}
\{x_{0i}'\} =\{x_{0i}\}+b_{i},\qquad
\{x_{0i}''\}=\{x_{0i}\}+2b_{i}-1,
\end{equation*}
or
\begin{equation*}
\{x_{0i}'\} =\{x_{0i}\}+b_{i}-1,\qquad
\{x_{0i}''\}=\{x_{0i}\}+2b_{i}-1,
\end{equation*}
or
 \begin{equation*}
 \{x_{0i}'\} =\{x_{0i}\}+b_{i}-1,\qquad
\{x_{0i}''\}=\{x_{0i}\}+2b_{i}-2.
\end{equation*}

 Let $I_{i1},I_{i2},I_{i3}$ be the intervals on the right-hand
 side of (\ref{int1}) or (\ref{int2}), depending on whether $b_{i}\leq
 \frac{1}{2}$ or not, respectively. Denote:
 $$\Omega_{\eta_{1} \eta_{2} ...\eta_{K-1}}=I_{1
\eta_{1}}\times I_{2 \eta_{2}}\times...\times
I_{K-1,\eta_{K-1}},\qquad \eta_{1},...,\eta_{K-1} \in \{1,2,3\}.$$
 The sets $\Omega_{ \eta_{1} \eta_{2} ...\eta_{K-1}}$ decompose the
$(K-1)$-dimensional torus into a union of
 $3^{K-1}$ disjoint boxes:
 $$\textbf{T}^{K-1}=
\bigcup_{\eta_{1}=1}^{3}\bigcup_{\eta_{2}=1}^{3}...\bigcup_{\eta_{K-1}=1}^{3}\Omega_{
\eta_{1} \eta_{2} ... \eta_{K-1}}.$$

  The information provided by the vector
$\{\x_{0}\}$ is partly redundant as the fact that $\sum_{i=1}^K
\{x_{0i}\}$ is an integer determines each component in terms of
the others. To avoid this inconvenience, we shall eliminate, say,
$\{x_{0K}\}$. Divide $\textbf{T}^{K-1}$ into $K$ parts as follows:
\begin{equation}\label{for12}
 \Omega^{s}=\{(t_{1},...,t_{K-1}) \in \textbf{T}^{K-1}:
s-1<\sum_{i=1}^{K-1} t_{i}\leq s\}, \quad
 s=0,...,K-1.\end{equation}
(Thus, $\Omega^{0}=\{\bf 0\}$, while all other $\Omega^{i}$'s have
non-empty interior.) Suppose that $(\{x_{01}\},...,\{x_{0,K-1}\})
\in \Omega^{s}.$ Then:
\begin{equation}\label{for2}
\{x_{0K}\}=s-\{x_{01}\}-...-\{x_{0,K-1}\}.
\end{equation}

We need a further subdivision to ensure that, in each cell, both
$\{x'_{0K}\}$ and $\{x''_{0K}\}$ assume the same form in
 terms of $\{x_{01}\},...,\{x_{0,K-1}\}.$ To this end, we first
 split \textbf{T} into three subintervals, similarly to
 (\ref{int1}) and (\ref{int2}), depending on whether $b_{K}\leq
 \frac{1}{2}$ or $b_{K}>\frac{1}{2}$, namely
\begin{equation}\label{int3}
\textbf{T}=[0,1-2b_{K})\cup[1-2b_{K},1-b_{K})\cup[1-b_{K},1)
\end{equation}
or
\begin{equation}\label{int4}
\textbf{T}=[0,1-b_{K})\cup[1-b_{K},2-2b_{K})\cup[2-2b_{K},1).
\end{equation}
Let $I_{K1}, I_{K2}, I_{K3}$ be the intervals in the splitting.
Put:
\begin{equation}\label{for3}
\Omega_{\eta}^{s}=\{(t_{1},...,t_{K-1}) \in \Omega^{s}:
s-\sum_{i=1}^{K-1} t_{i} \in I_{K \eta}\}, \qquad \eta=1,2,3.
\end{equation}

Suppose $b_{K}\leq\frac{1}{2}$. If the point
$(\{x_{01}\},...,\{x_{0,K-1}\}) $ belongs to $ \Omega_{1}^{s}$,
then

 \begin{equation*}\begin{array}{l}
\{x'_{0K}\}=s-\sum_{i=1}^{K-1}\{x_{0i}\} +b_{K}, \qquad
\{x''_{0K}\}=s-\sum_{i=1}^{K-1}\{x_{0i}\}+2b_{K},
\end{array}
\end{equation*}
if it belongs to $ \Omega_{2}^{s}$, then
\begin{equation*}\begin{array}{l}
\{x'_{0K}\}=s-\sum_{i=1}^{K-1}\{x_{0i}\}+b_{K}, \qquad
\{x''_{0K}\}=s-\sum_{i=1}^{K-1}\{x_{0i}\}+2b_{K}-1,
\end{array}
\end{equation*}
and if it belongs to $ \Omega_{3}^{s}$, then
\begin{equation*}\begin{array}{l}
\{x'_{0K}\}=s-\sum_{i=1}^{K-1}\{x_{0i}\}+b_{K}-1, \qquad
\{x''_{0K}\}=s-\sum_{i=1}^{K-1}\{x_{0i}\}+2b_{K}-1.
\end{array}
\end{equation*}
\noindent If $ b_{K}> \frac{1}{2},$ then we similarly find linear
expressions for $\{x'_{0K}\}$ and $ \{x''_{0K}\}$ in terms of the
$\{x_{0i}\}$'s on each $\Omega_{\eta}^{s}$.

Denote:
 $$\Omega_{ \eta_{1} \eta_{2} ... \eta_{K}}^{s}=
 \Omega_{ \eta_{1} \eta_{2} ... \eta_{K-1}} \cap \Omega_{\eta_K}^{s} ,
\qquad  0\leq s\leq K-1,\ 1\leq \eta_i\leq 3.$$ Then
$$\textbf{T}^{K-1}=\bigcup_{s=0}^{K-1}\bigcup_{\eta_{1}=1}^{3}...
\bigcup_{\eta_{K}=1}^{3} \Omega_{ \eta_{1} ... \eta_{K}}^{s}$$
forms a decomposition of $\textbf{T}^{K-1}$ into a disjoint union
of $K\cdot3^{K}$ disjoint polytopes. The important property of
this decomposition  is that, if $(\{x_{01}\},...,$ $
\{x_{0,K-1}\})$ belongs to any cell $\Omega_{ \eta_{1} \eta_{2}
... \eta_{K}}^{s}$, the $2K+1$ numbers $ \{x_{0K}\}, \{x_{01}'\},
..., $ $ \{x_{0K}'\},\{x_{01}''\},...,$ $ \{x_{0K}''\}$ depend
linearly on the first $K-1$ coordinates $\{x_{01}\},...,$ $
\{x_{0,K-1}\}$. That is

\begin{equation}\begin{aligned}\label{for4}
&\{x'_{0i}\}=\{x_{0i}\}+b_{i}-\alpha'_{i},\qquad &1\leq i \leq K-1,\\
&\{x''_{0i}\}=\{x_{0i}\}+2b_{i}-\alpha''_{i},\qquad &1\leq i \leq
K-1,
\end{aligned}\end{equation}
\begin{equation}\begin{aligned}\label{for5}
&\{x'_{0K}\}=s-\sum_{i=1}^{K-1}\{x_{0i}\}+b_{K}-\alpha'_{K},\qquad\\
&\{x''_{0K}\}=s-\sum_{i=1}^{K-1}\{x_{0i}\}+2b_{K}-\alpha''_{K},
\end{aligned}\end{equation}
 where $\alpha'_{i}\in \{0,1\},\
\alpha''_{i}\in \{0,1,2\}$, $i=1,2,...,K$. Altogether, there
exists a linear transformation $ T:\textbf{R}^{K-1}\rightarrow
\textbf{R}^{K}$, and  for each cell $\Omega_{ \eta_{1} \eta_{2}
... \eta_{K}}^{s}$ there exist vectors $\v, \v', \v'' \in
\textbf{R}^{K}$, such that, denoting
$\x=(\{x_{01}\},...,\{x_{0,K-1}\})$, we have
\begin{equation}\label{for6}
 \{\x_{0}\}=T\x+\v,\quad \{\x'_{0}\}=T\x+\v',\quad \{\x''_{0}\}=T\x+\v''.
 \end{equation}
On each cell $\Omega_{ \eta_{1} \eta_{2} ... \eta_{K}}^{s}$ we may
now rewrite the system (\ref{doman1}), defining the optimal
vectors $\p, \p', \p''$, in the form:
\begin{equation}\label{domain2}\left\{\begin{aligned}
&2(\l-\p)^{t}W T \x\leq \l^{t}W\l-\p^{t}W\p-2(\l-\p)^{t}W\v, \ \
&\l &\in
L,\\
&2(\l-\p')^{t}W T \x\leq
 \l^{t}W\l-(\p')^{t}W\p'-2(\l-\p')^{t}W\v', \ \ &\l &\in L',\\
&2( \l-\p'')^{t}W T \x\leq
\l^{t}W\l-(\p'')^{t}W\p''-2(\l-\p'')^{t}W\v'', \ \ &\l &\in L''.
\end{aligned} \right.\end{equation}
Note that we have suppressed the dependence of the sets $L, L',
L''$ on $\x_{0}, \x'_{0}, \x''_{0}.$ In fact, considering $L$, for
example, it is clear that each candidate $l \in L$ must have sum
of coordinates $s$ and, in view of Lemma \ref{lem3}, its norm is
bounded above by $\Delta+\sqrt{K}$. Thus, taking
\begin{equation*}\begin{aligned}
L&=\{\l \in \textbf{Z}^{K}: \|\l\|\ \leq \Delta+\sqrt{K},\ \
\sum_{i=1}^{K}l_{i}=s \},\\
 L'&=\{\l \in
\textbf{Z}^{K}: \|\l\|\ \leq \Delta+\sqrt{K},\ \
\sum_{i=1}^{K}l_{i}=s+1-\sum_{i=1}^{K}\alpha'_{i} \},\\
 L''&=\{\l
\in \textbf{Z}^{K}: \|\l\|\ \leq \Delta+\sqrt{K},\ \
\sum_{i=1}^{K}l_{i}=s+2-\sum_{i=1}^{K}\alpha''_{i}
\},\end{aligned} \end{equation*} (where
$\l=(l_{1},l_{2},...,l_{K})$), we certainly do not miss any
potentially optimal vectors $\p, \p', \p''$ by restricting the
search to $L, L', L''$, respectively.

 For each choice of $\eta_{1}, ...,
\eta_{K},s$ and of the vectors $\p,\p',\p''$, let $P_{ \eta_{1}
... \eta_{K}}^{s\p\p'\p''}$ be the set of all points in $\Omega_{
\eta_{1} ... \eta_{K}}^{s}$ satisfying (\ref{domain2}). Then
\begin{equation} \label{torus}
\textbf{T}^{K-1}=\bigcup_{s=0}^{K-1}\bigcup_{\eta_{1}=1}^{3}...
\bigcup_{\eta_{K}=1}^{3} \bigcup_{\p \in L}\bigcup_{\p' \in
L'}\bigcup_{\p'' \in L''} P_{ \eta_{1} ...
\eta_{K}}^{s\p\p'\p''},\end{equation} where the sets on the
right-hand side are disjoint (up to sets of a smaller dimension).

By (\ref{vec}) and (\ref{for6}), for all points in each
subpolytope $P_{ \eta_{1} ... \eta_{K}}^{s\p\p'\p''}$  we have the
same optimal correction vectors
 \begin{equation} \label{domain3}
   \y_{1}= \q-T\x,\quad
   \y_{2}= \q'-T\x,\quad
   \y_{3}=\q''-T\x,
 \end{equation}
 where $\q=\p-\v, \q'=\p'-\v', \q''=\p''-\v''.$ Hence, if the point $\x$
 belongs to $ P_{ \eta_{1} ...\eta_{K}}^{s\p\p'\p''}$, then $\chi_{N}$
 depends linearly on the coordinates $\{x_{01}\},..., \{x_{0,K-1}\}$ :
 \begin{equation}\begin{aligned}\label{chi1}
 \chi_{N}= 1+&(\q-T\x)^tW(\q-T\x)
-2(\q'-T\x)^tW(\q-T\x)+\\
+(&\q''-T\x)^tW(\q''-T\x) \\
=2(-&\q+2\q'-\q'')^tW T\x +const,
 \end{aligned}\end{equation}
where $const=1+\q^tW\q-2(\q')^tW\q'+(\q'')^tW\q''$.(Note that all
coefficients on the right-hand side (\ref{chi1}) depend on
$\eta_{1}, ...,\eta_{K},s, \p, \p', \p''$.)

 We need to find the function $F$ according to which the sequence
 $(\chi_{N})_{N=1}^{\infty}$ is asymptotically $F$-distributed. To
simplify our notations, rewrite (\ref{torus}) in the form
$$\textbf{T}^{K-1}=\bigcup_{i=1}^{r}P_{i},$$
where each $P_{i}$ is one of the polytopes $P_{ \eta_{1}
...\eta_{K}}^{s\p\p'\p''}.$ Denote
 $$\bftheta_{N}=(\{Nb_{1}\},...,\{Nb_{K-1}\}), \qquad N=1,2,...\textbf{,}$$

\noindent and
$$A_{i}=\{{N \in \textbf{N}}: \bftheta_{N} \in P_{i}\}, \qquad i=1,2,...,r.$$

\noindent(The $A_{i}$'s may intersect, as the $P_{i}$'s may
intersect on sets of a smaller dimension. However, this will cause
no problem as the intersections are sets of density $0$ in
$\textbf{N}.$ Alternatively, we may first decide in some arbitrary
way where to place ``problematic'' integers.)

 Let $\bftheta_{N}^{(i)}$ be the subsequence of
$\bftheta_{N}$, consisting of those elements $\bftheta_{N}$ with
$N \in A_{i}$, $1 \leq i \leq r $. That is, $\bftheta_{N}^{(i)}$
is the $N$-th element of $(\bftheta_{N})_{N=1}^{\infty}$ which
belongs to $A_{i}$. Let $(\chi_{N}^{(i)})_{N=1}^{\infty}$ be the
corresponding subsequence of $(\chi_{N})_{N=1}^{\infty}$. If the
point $\x$ lies in $ P_{i}$, then it belongs to the subsequence
$(\bftheta_{N}^{(i)})_{N=1}^{\infty}$. Thus, by (\ref{chi1}) there
are exist affine functions $\psi_{i}: \bf{R}\rm
^{K-1}\rightarrow\bf{R}$ such that
 \begin{equation}\label{chi2}
 \chi_{N}^{(i)}=\psi_{i}(\bftheta_{N}^{(i)}), \qquad 1\leq i\leq r,\ \ N=1,2,...\textbf{.}
 \end{equation}

 Since the numbers $b_{1},...,b_{K}$ are linearly independent over
 \textbf{Q}, so are the numbers $1,b_{1},...,b_{K-1}$, and consequently
 the sequence $(\bftheta_{N})$ is uniformly distributed modulo $1$ in
 $\textbf{R}^{K-1}$. Hence, each of the subsequences
$(\bftheta_{N}^{(i)})_{N=1}^{\infty},\ \  1\leq i\leq r$, is
uniformly distributed in $P_{i}.$ By (\ref{chi2}) the sequence
$(\chi_{N}^{(i)})_{N=1}^{\infty}$ is the image of a uniformly
distributed sequence in $P_{i}$ under the mapping $\psi_{i}.$
Hence, letting $ (Y_{1},Y_{2},...,Y_{K-1})$ be a
$(K-1)$-dimensional random variable, uniformly distributed in
$P_{i}$, we see that $(\chi_{N}^{(i)})_{N=1}^{\infty}$ is
$F_{i}$-distributed, where $F_{i}$ is the distribution function of
$ \psi_{i}(Y_{1},Y_{2},...,Y_{K-1})$.

 According to \cite[Thm 2.3]{Berend}, $F_{i}$ is a piecewise polynomial
 function, each polynomial piece being of degree at most $K-1$, and
 can be effectively computed.
Since  $(\bftheta_{N}^{(i)})_{N=1}^{\infty}$ is
 uniformly distributed modulo $1$ in $\textbf{R}^{K-1}$, the density
 of each $A_{i}$ is the measure $d_{i}$ of the set $P_{i}$. As $P_{i}$
 is a polytope, this measure can be effectively computed. By \cite[Lemma 1]{ABB1},
$(\chi_{N})_{N=1}^{\infty}$ is asymptotically $F$-distributed,
where $F=\sum_{i=1}^{r}d_{i}F_{i}$. This completes the proof.


\newpage
\begin{thebibliography}{1}
\bibitem{Ashoori} R. C. Ashoori, H. L. Stormer, J. S. Weiner, L. N. Pfeiffer, S. J. Pearton, K. W. Baldwin, and K. W. West
 {\it Phys. Rev. Lett.}  68, (1992) 3088-3091,
R. C. Ashoori {\it et al.} , {\it Physica}, Amsterdam, 189B, (1993) 117;
D. Berman, N. B. Zhitenev, R. C. Ashoori, and M. Shayegan, Observation of Quantum Fluctuations of Charge on a Quantum Dot, {\it Phys. Rev. Lett},
\textbf{82}(1999), 161--164.
\bibitem{ABB1}Y. Avishai, D. Berend and R. Berkovits, Statistics of
addition spectra of independent quantum systems, {\it Jour. of
Phys.} A (Math. Gen.) 31 (1998), 8063-8072.

\bibitem{Berend} D. Berend and L. Bromberg, Uniform decompositions of
polytopes, preprint.

%
\bibitem{Landau}  L.D.Landau, I.M.Livshiz {\it Electrodynamics of
Continuous Media} (Pergamon, 1959). See, especially discussion in
pages 3-7.

\bibitem{KN} L. Kuipers and H. Niederreiter,
{\it Uniform Distribution of Sequences}, Wiley, New York, 1974.

%
\end{thebibliography}
\end{document}